\documentclass[%
 reprint,
superscriptaddress,
 amsmath,amssymb,
 aps,
]{revtex4-2}

\usepackage{graphicx}
\usepackage{dcolumn}
\usepackage{bm}

\begin{document}

\preprint{APS/123-QED}

\title{High repetition rate Time-Resolved  VUV ARPES at 10.8 eV photon energy}

\author{Simone Peli}
\email{simone.peli@elettra.eu}
 \affiliation{Elettra - Sincrotrone Trieste S.C.p.A., Strada Statale 14, km 163.5, Trieste, Italy}
 \author{Denny Puntel}
 \affiliation{Elettra - Sincrotrone Trieste S.C.p.A., Strada Statale 14, km 163.5, Trieste, Italy}
 \author{Damir Kopic}
 \affiliation{Elettra - Sincrotrone Trieste S.C.p.A., Strada Statale 14, km 163.5, Trieste, Italy}
 \author{Benjamin Sockol}
 \affiliation{Department of Physics, Massachusetts Institute of Technology, Cambridge, MA 02139, USA}
 \author{Fulvio Parmigiani}
 \affiliation{Elettra - Sincrotrone Trieste S.C.p.A., Strada Statale 14, km 163.5, Trieste, Italy}
\author{Federico Cilento}%
 \email{federico.cilento@elettra.eu}
\affiliation{Elettra - Sincrotrone Trieste S.C.p.A., Strada Statale 14, km 163.5, Trieste, Italy}%





\begin{abstract}
The quest for mapping the femtosecond dynamics of the electronic band structure of complex materials via Time- and Angle-Resolved Photoelectron Spectroscopy (TR-ARPES) over their full First Brillouin Zone is pushing the development of schemes to efficiently generate ultrashort photon pulses in the VUV-range of photon energies. At present, the critical aspect is to combine a high photon energy with high photoemission count rates and a small pulse-bandwidth, necessary to achieve high energy resolution in ARPES, while preserving a good time resolution and mitigating space-charge effects. Here we describe a novel approach to produce light pulses at 10.8 eV, combining high repetition rate operation (1-4 MHz), high energy resolution ($\sim26$ meV) and space-charge free operation, with a time-resolution of $\sim$700 fs. These results have been achieved by generating the 9th harmonics of a Yb fiber laser, through a phase-matched process of third harmonics generation in Xenon of the laser third harmonics. The full up-conversion process is driven by a seed pulse energy as low as 10 $\mu$J, hence is easily scalable to multi-MHz operation. This source opens the way to TR-ARPES experiments for the investigation of the electron dynamics over the full first Brillouin zone of most complex materials, with unprecedented energy and momentum resolutions and high count rates. The performances of our setup are tested in a number of experiments on WTe$_2$ and Bi$_2$Se$_3$, of which we measure the electronic band structure in energy, two-dimensional momentum and time.
\end{abstract}

\maketitle


\section{\label{sec:Introduction}Introduction}
The quantities that describe the electrons in a solid are binding energy ($E_B$), momentum ($k$) and spin ($s$). In this frame, the Angle-Resolved Photoemission Spectroscopy (ARPES) has gained a prominent role for its ability of directly measuring the energy and the momentum of the electrons simultaneously \cite{Smith1973}. Spin-resolved ARPES adds the information about the spin polarization, to achieve a complete overview on the electronic structure of the material \cite{Hsieh2009}. Thanks to a continuous technological improvement of the electron analyzers and the developments of new photon sources, ARPES has matured to become the leading technique in the world of condensed matter physics \cite{Damascelli2003, Zhou2007}.

ARPES is based on the photoelectric effect for which, by means of an incident photon beam, the electrons are extracted out of materials with a defined kinetic energy $E$ and emission angle $\theta$. The extracted electrons are then collected by a hemispherical analyzer that distributes each electron on a two-dimensional map with $E$ and $\theta$ axes. Knowing the kinetic energy and the emission angle of the electron allows to retrieve its binding energy $E_{B}$ and its  parallel momentum $\hbar k_{\|}$ inside the material thanks to the energy and momentum conservation laws for the photoemission process:
\begin{eqnarray}
E=h\nu-W-E_{B}
\label{Energy_Conservation}
\\
\hbar k_{\|}=\sqrt{2m_{e}E}\cdot \sin \theta
\label{Momentum_Conservation}
\end{eqnarray}
where $h \nu$ is the incident photon energy, $W$ is the material work function and $m_{e}$ is the electron mass. Eq.\ref{Energy_Conservation} indicates that higher photon energies enable to access higher binding energies. Eq.\ref{Momentum_Conservation} states that the accessible momenta are restricted by the kinetic energy which, again, is limited by the photon energy. Hence the accessible ranges of the fundamental quantities measured by photoemission are both determined by the photon energy, which turns out to be a key parameter in the ARPES experiment.

Besides the traditional rare gas discharge lamps that provides fixed photon energies \cite{Harter2012}, in the last three decades the synchrotron light sources have become the leading facilities for studying the electronic structure of materials. Synchrotron light can provide photon energy ranging from the infrared to the soft and hard X-ray region with high brilliance, a full control on the polarization of light and on the beam size on the sample. However, it has some drawbacks: (1) it's difficult to achieve a very high energy resolution ($< 2$ meV) maintaining a reasonable photon flux for the experiment; (2) the pulse duration (several tens of picoseconds) is not suitable to access the electronic dynamics. To overcome these limitations, new ARPES endstations based on table-top laser light sources have emerged in the last years \cite{Zhou2018}. The use of laser light with a very high intensity confined in very short pulses (typically $< 1$ ps) allows to exploit non linear effects in crystals and gases to generate, starting from the fundamental near-infrared wavelength, photon energies that are suitable for ARPES experiments. Moreover, the use of short pulses allows the realization of pump-probe experiments to directly observe the bands relaxation in a time-resolved ARPES frame.

In table-top laser-based ARPES the ultraviolet (UV) and vacuum-ultraviolet (VUV) light required for the photoemission is commonly generated with two methods. The first method is a non-linear frequency up-conversion achieved in non-linear optical crystals, where the fundamental light, through processes of second harmonic generation (SHG) and sum-frequency generation (SFG), is converted in light with higher photon energy. The second method is the high harmonic generation (HHG) process \cite{Ferray1988} that exploits the ionization of noble gas atoms and the subsequent recombination to generate extreme ultra-violet (EUV) light. Besides these table-top techniques there are free electron lasers (FELs) which are able to generate fs pulses in a tunable range (10-1000 eV) with a very high flux ($10^{13-14}$ photons per pulse) but very low repetition rate (50 Hz).

The frequency up-conversion in non-centro symmetric crystals is the most used method for time-resolved ARPES with high-resolution. The main limitation is given by the transparency of the crystal employed, that restricts the maximum photon energy achievable. For example, the absorption edge of a BBO (Beta Barium Borate) crystal is 6.56 eV \cite{Liu2008} and the maximum photon energy achievable with a SHG process in phase-matching condition is 6.05 eV. Although the novel KBBF crystal can push the limit of frequency conversion up to 7.56 eV, this remains the maximum photon energy reachable with non-linear optical processes in crystals \cite{Zhou2018}. With this photon energy, taking a reasonable $45^{\circ}$ sample rotation, the maximum achievable momentum is 0.77 \AA$^{-1}$ which is quite far from the edge of the first Brillouin zone of most quantum materials, positioned at $>1$ \AA$^{-1}$. This limitation is overcome using an HHG process that starts with the illumination of a noble gas by infrared or visible light of very high intensity ($10^{14} \mathrm{\text{ } W\cdot cm^{-2}}$). Such intense electric field ionize the gas atoms making the electrons free. These electrons are then accelerated by the laser light electric field, gaining kinetic energy. When the electrons are pushed again in the vicinity of the ion, they can recombine emitting a photon at high energy. The photon energy generated is limited only by the kinetic energy accumulated after the ionization and before the recombination and normally is of the order of 15-40 eV. Since the recombination probability is  of the order of $10^{-6}$, the efficiency of the frequency conversion is very low. This low efficiency require a very strong intensity per pulse in order to obtain a reasonable photon flux and this limits the operation repetition rate at $<250$ KHz \cite{Cucini2019}. Since HHG is a highly nonlinear process, it is strongly dependent on a number of experimental parameters. In addition to the laser wavelength, the pulse energy and duration, it is necessary to optimize the beam profile, the focusing geometry, the interaction length, and the gas pressure, making this technique challenging from a technical and engineering point of view \cite{Frietsch2013}.

We propose here a different approach for time-resolved ARPES, designed to overcome the limited accessibility to the Brillouin zone of the non-linear-crystals-based-techniques and the low throughput of HHG based systems. This technique leads to VUV pulses with 10.8 eV photon energy, with a tunable repetition rate up to 4 MHz and a time resolution of 700 fs. The VUV light is the result of a third harmonic generation (THG) process in Xenon gas as further discussed later. Bernsten et al. \cite{Bernsten2011} conducted the first ARPES experiment using the THG process in Xenon to generate 10.5 eV pulses at 8 MHz repetition rate with a pulse length of 10 ps. Only many years later, He et al. \cite{He2016} developed a system to generate 100 ps pulses at 10.9 eV of photon energy with a repetition rate of 10 MHz exploiting a four wave mixing SFG in Xenon. With this source they performed equilibrium ARPES experiments on high-temperature cuprates superconductors and iron-based superconductors. In 2017 Zhao et al. \cite{Zhao2017} realized a system to produce pulses at 10.7 eV with a repetition rate of 1 MHz with a nominal pulse duration of 100 fs from a THG process in a mixture of Xenon and Argon gases. However, this setup was not coupled to an ARPES apparatus. As per our knowledge, the only time-resolved ARPES experiment with VUV laser light at $\sim 11$ eV reported so far is by Zong et al. \cite{Zong2019}. They studied the charge density wave system LaTe3 with VUV pulses at 10.75 eV generated through THG in a hollow fiber filled with Xenon gas. The system works at 250 KHz and has a temporal resolution of 230 fs. In this panorama, our time-resolved VUV ARPES experimental setup ranks in the first place in terms of throughput, photon energy, energy resolution, stability and reliability and in one of the first places in terms of time resolution. In this paper, after the description and the details of the experimental setup we take the Bi$_2$Se$_3$ topological insulator and the WTe$_2$ semimetal as case studies. We measured the photoemission intensity I($E_B$, $k_x$, $k_y$, $t$) including the unoccupied states. This system put together the advantages of a non-linear efficient light conversion in gas with a photon energy that allows to investigate the full Brillouin zone of most materials and more bounded electronic states.

\section{\label{sec:Experimental_Setup}Experimental setup}
\subsection{\label{sec:Generation}Generation of 10.8 eV light}
\begin{figure}
\includegraphics[width=\columnwidth]{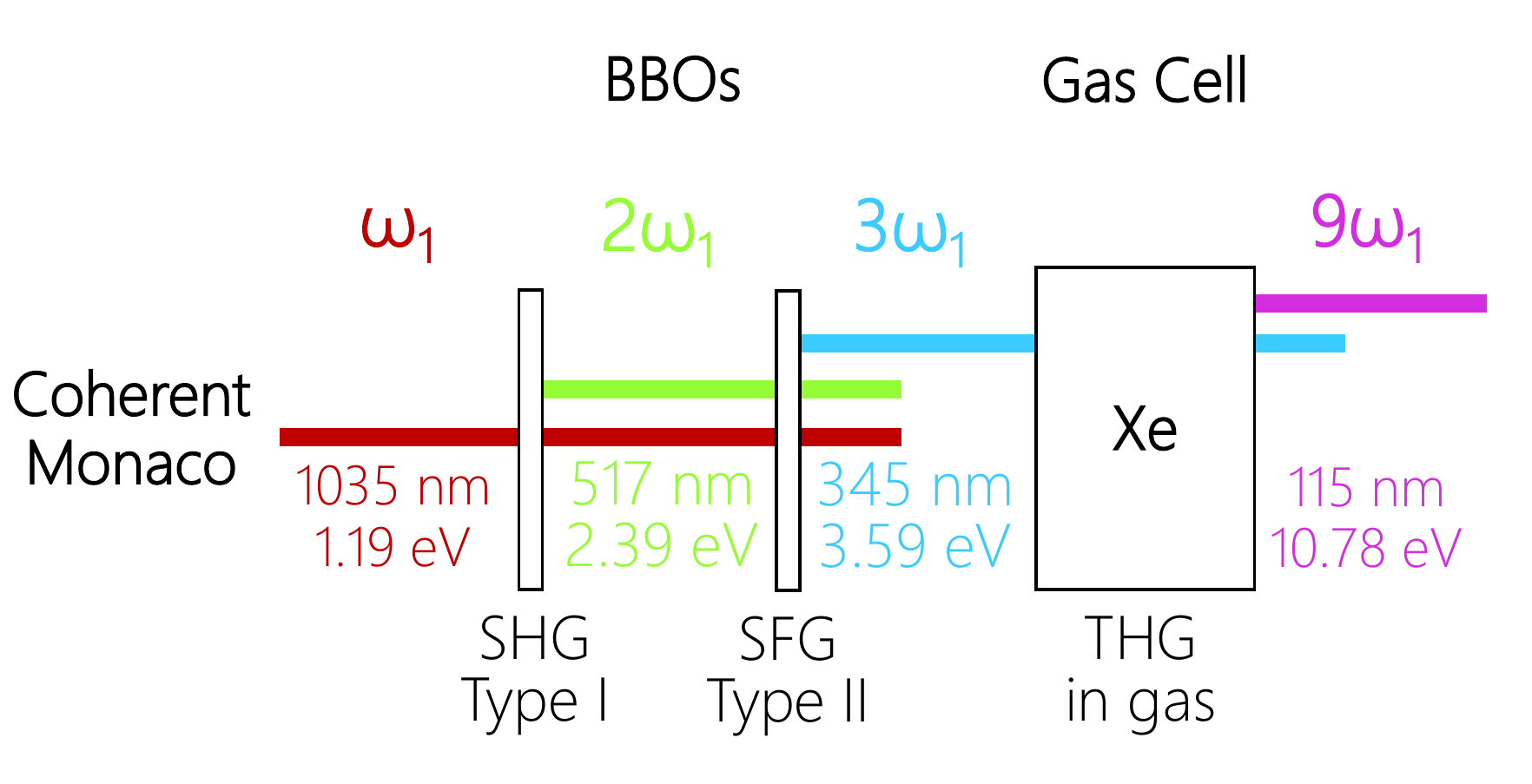}
\caption{\label{fig:Frequency_Conversion_Stages}\textbf{Frequency up-conversion stages for VUV light generation}. The fundamental light at frequency $\omega_1$ from the Coherent Monaco laser is frequency-doubled in a type-I BBO crystal. Second harmonic $\omega_2$ and fundamental $\omega_1$ are frequency-summed in a type-II BBO crystal to generate the third harmonic $\omega_3$ at 345 nm. Finally, a THG process in Xenon gas is used to produce the VUV ninth harmonic $\omega_9$ at 115nm (10.8 eV).}
\end{figure}

The 10.8 eV light is the result of three stages of frequency up-conversion. Frequency conversion occurs in non-linear media after the interaction with high intensity electric fields. In these conditions the polarization $P$ of the medium has a non-linear response to the electric field $E$ of the light:
\begin{eqnarray}
P(t)=\epsilon_0(\chi^{(1)}E(t)+\chi^{(2)}E^2(t)+\chi^{(3)}E^3(t)+\dots)
\label{NonLinear_Polarization}
\end{eqnarray}
where $\chi^{(n)}$ are the $n$-th order susceptibility of the medium and $\epsilon_0$ the vacuum permittivity. For symmetry reasons \cite{Reintjes1984}, in a non-centrosymmetric crystal, the second order term in Eq.\ref{NonLinear_Polarization} is non-zero so that second order effects such as SHG and SFG can occur. For similar arguments, in gases, which are centrosymmetric, the second order susceptibility vanishes and the first non-linear term is the third order one ($\chi^{(3)}$) that leads to third harmonic generation (THG). The phase matching condition for frequency up-conversion in gases takes advantage from the anomalous dispersion that accompanies the allowed dipole transitions. In particular, anomalous dispersion is Xenon occurs in the spectral region between 113.5 nm (10.92 eV) and 117 nm (10.59 nm) \cite{Ganeev1996, Ganeev2000}. When the third harmonic falls in this range, \emph{i.e.} the driving wavelength is in the 340.5-351 nm wavelength range, an efficient phase matched THG can occur \cite{Kung1973}. Nowadays high-power and high-repetition rate Yb:fiber-based lasers are very common in optics laboratories. Such systems have an output wavelength of $\omega_1\simeq 1035$ nm and, by means of standard BBO crystals, it is possible to obtain its third harmonic at $\omega_3=345$ nm, that is a suitable driving wavelength for phase-matched THG in Xenon (see Figure \ref{fig:Frequency_Conversion_Stages}). Hence, starting from laser systems with output at 1035 nm, it is possible to generate the VUV ninth harmonic ($\omega_9$) of the fundamental wavelength with an efficiency of the order of $10^{-4}$-$10^{-5}$ \cite{Zhao2017}.

\subsection{\label{sec:Experimental_Layout}Experimental Layout}
The fundamental light used to generate the VUV 10.8 eV photons is the output at 1035 nm (1.19 eV) of a \textit{Coherent Monaco 1035}, a femtosecond Yb:fiber-based laser that generates pulses of 290 fs and energy of $40$ $\mu$J, with a base repetition rate (r.r) of 1 MHz and 40 W of power. In this condition, the r.r. can be set by a pulse-picker, down to single-shot operation. Alternatively, the r.r. can be increased up to 50 MHz, at the expense of the energy/pulse.

\begin{figure}
\includegraphics[width=\linewidth]{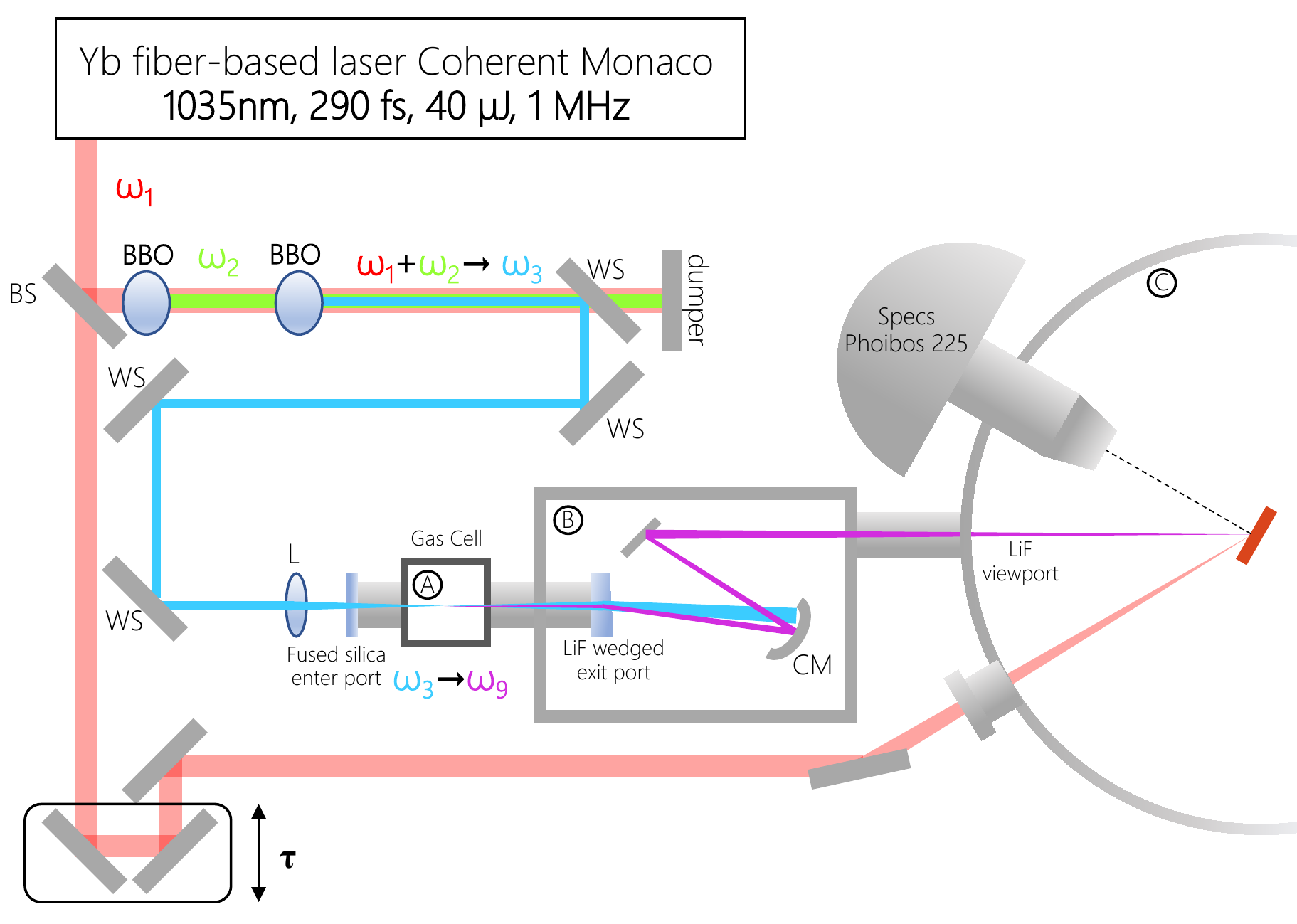}
\caption{\label{fig:Experimental_Setup}\textbf{Layout of the experimental setup.} The fundamental light at frequency $\omega_1$ is split in pump and probe by a beam splitter (BS). The pump goes on the sample after passing through a mechanical delay stage. The probe beam undergoes two frequency up-conversion stages in BBO crystals. The third harmonic is selected among the remaining components through a series of wavelength separators (WS). The THG process for ninth harmonic ($\omega_9$) generation takes place in a small cell filled by Xenon gas (ambience A). The separation of the $\omega_3$ and $\omega_9$ beams occurs in a low vacuum refocusing chamber (ambience B) where the VUV light is then focalized by a curved mirror (CM) and driven on the sample in the UHV photoemission chamber (ambience C) through a LiF window.}
\end{figure}

Following the scheme of Figure \ref{fig:Experimental_Setup} the fundamental light $\omega_1$ first undergoes a frequency doubling in a 2.5 mm thick type-I BBO crystal, cut at an angle of $\theta=23.3^{\circ}$, to generate photons at 517 nm (2.39 eV) with a maximum efficiency of 50\%. The beams at $\omega_1$ and $\omega_2$ enter collinearly in a type-II BBO crystal for a SFG process to generate the third harmonic $\omega_3$ of the fundamental at 345 nm (3.59 eV). The thickness of this crystal is 1.7 mm and the cut angle is $\theta=32.3^{\circ}$. The total maximum efficiency of third harmonic generation is 20\%. The choice of a type-II crystal simplifies the design since the pulses duration is $\simeq 300$ fs. Hence, the time separation between the fundamental and the second harmonic accumulated in the first crystal is negligible, thus avoiding any need for delay compensation. The third harmonic $\omega_3$ at 345 nm is then cleaned from the lower harmonics through a series of wavelength separators reflecting 345 nm light. The THG beam is then driven on a fused silica lens (L) of 100 mm focal length that focuses the light in a small cell (ambience A) filled with Xenon gas (at a pressure of 160 mbar). Here, through a four-wave mixing THG process, the ninth harmonic $\omega_9$ at 115 nm (10.78 eV) is generated. After the generation, the VUV light must travel either in vacuum or nitrogen purged ambiences because its wavelength is absorbed by the air components. The isolation between the gas cell and the next refocusing chamber (ambience B) is guaranteed by a LiF wedged exit port which has an absorption threshold at around 120 nm and has a transmittivity of $\sim20\%$ at 115 nm. The wedged window has an apex angle of $5^{\circ}$ and works as a dispersing prism to separate the ninth harmonic from the third harmonic. The refocusing chamber is purged with a constant flux of nitrogen gas and it is pumped by a scroll pump to keep a constant pressure of 30 mbar. This flux allows to maintain the temperature of the exit LiF port under control avoiding any heating effect that would affect the transmittivity of the window itself. After the separation the VUV light at 10.8 eV is reflected back from a curved mirror with a radius of curvature of 250 mm that has been chosen to focalize the beam on the sample. A motorized plane mirror finally drives the light inside the ARPES chamber (ambience C). Both the curved and the plane mirrors in the refocusing chamber have a specific coating for VUV light. At the entrance of the ARPES chamber a LiF viewport allows the refocusing chamber to be separated by the ultra-high vacuum ambience dramatically simplifying the design of the system. The VUV light hits the sample with an incidence angle $\theta=30^{\circ}$ with respect to the sample normal. The hemispherical analyzer is a SPECS PHOIBOS with an energy resolution of $\sim 26$ meV. The sample is mounted on a six degrees-of-freedom manipulator with $x$, $y$, $z$ axis and polar angle $\theta$, azimuthal angle $\psi$ and tilt angle $\phi$. A polarizing beamsplitter plate is used to obtain the pump beam, whose power can be tuned by a lambda/2 waveplate. On the pump path, a mechanical translational stage controls the time delay $t$ between the pump and the probe pulses.

\section{\label{sec:Characterization}Characterization of the 10.8 eV beam}
\subsection{\label{sec:Spot_Size}Spot Size and Pressure Dependence}
\begin{figure}
\includegraphics[width=\columnwidth]{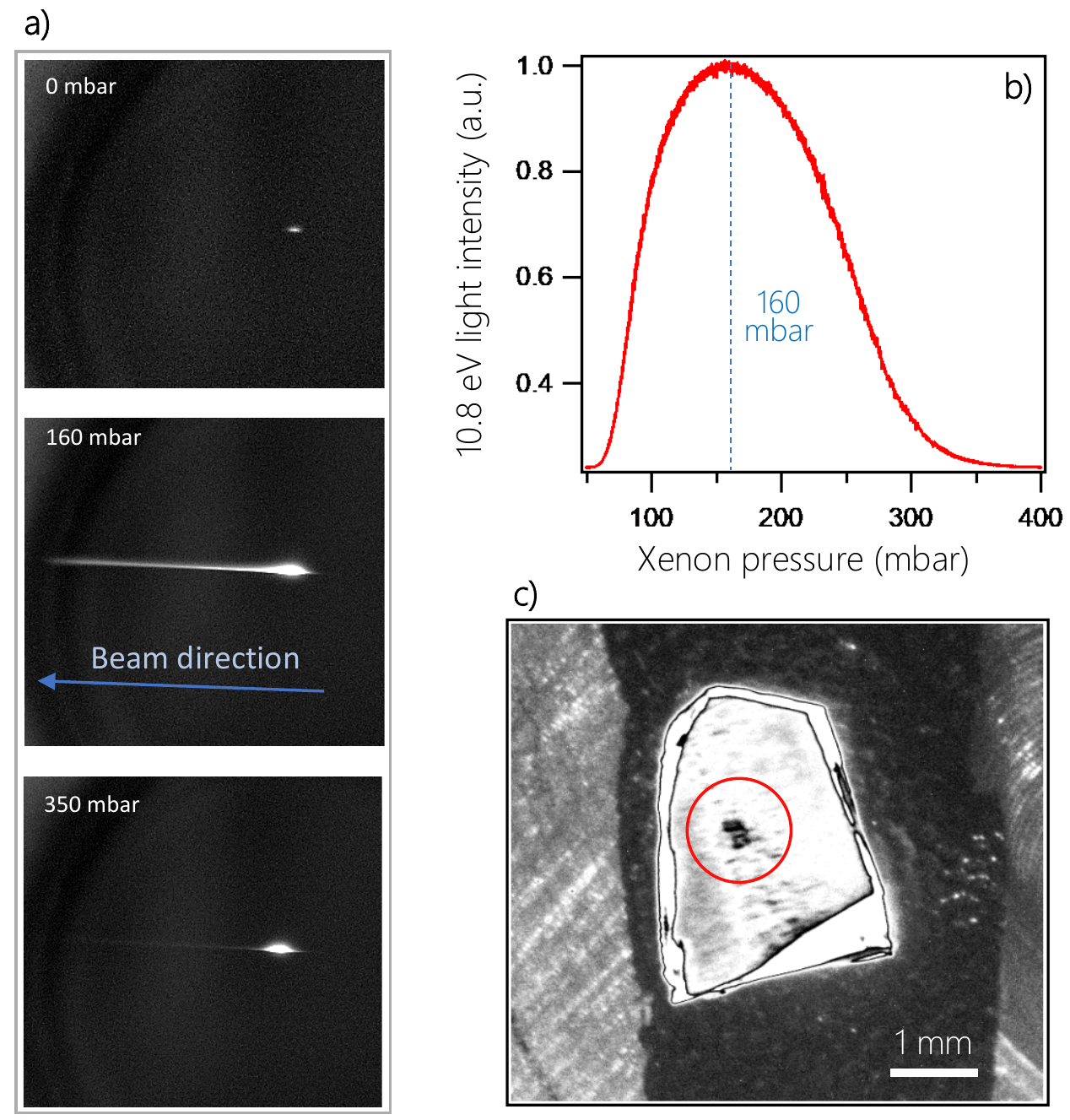}
\caption{\label{fig:Characteristics}\textbf{Characteristics of the VUV light.} The THG process in the Xenon gas cell is accompanied by a visible plasma whose intensity is changing with the Xenon pressure (\textbf{a}). The THG phase matching can be tuned varying the Xenon pressure in the cell leading to a maximum at $\simeq 160$ mbar (\textbf{b}). Panel (\textbf{c}) shows a negative photograph of the VUV light at 10.8 eV at the sample position on a YAG scintillator. The black area in the center of the red circle is the visible footprint of the focused VUV light. The red circle has a 1 mm diameter. The analysis of the photograph returns an average VUV spot diameter of $\simeq 260$ $\mathrm{\mu m}$.}
\end{figure}

The negative dispersion of Xenon in specific energy regions allows to have an efficient THG because of favorable phase-matching conditions in the tight focusing regime. The phase mismatch can be optimized by tuning the pressure of Xenon in the gas cell (A). We introduce the Xenon gas in the cell through a clean gas line and we monitor the gas pressure with a piezo pressure gauge. We recorded the 10.8 eV light intensity using a channeltron electron multiplier, while gradually increasing the gas pressure. The result is shown in Figure \ref{fig:Characteristics}(b).
We obtain the maximum generation at $\simeq 160$ mbar. When the beam at $\omega_3$ is focused in the gas cell, a bright plasma is observed (Figure \ref{fig:Characteristics}(a)); its brightness follows the same behaviour of the 10.8 eV light flux while changing the Xenon pressure. In correspondence of its maximum, a bright strip of plasma following the same direction of the beam is clearly visible. After the generation, the 10.8 eV beam is separated from the $\omega_3$ beam and is focused inside the photoemission chamber. In order to estimate the probe beam size, we placed a YAG scintillator in the position of the sample and we took a picture of the fluorescence with a CMOS camera (Figure \ref{fig:Characteristics}(c)). The spot shape is not perfectly circular probably because of the use of a spherical mirror not at the normal incidence. The vertical dimension is $\sim 350$ $\mu$m and the horizontal one is $\sim 170$ $\mu$m.

\subsection{\label{sec:Space_Charge}Space Charge and Energy Resolution}
\begin{figure}
\includegraphics[width=\columnwidth]{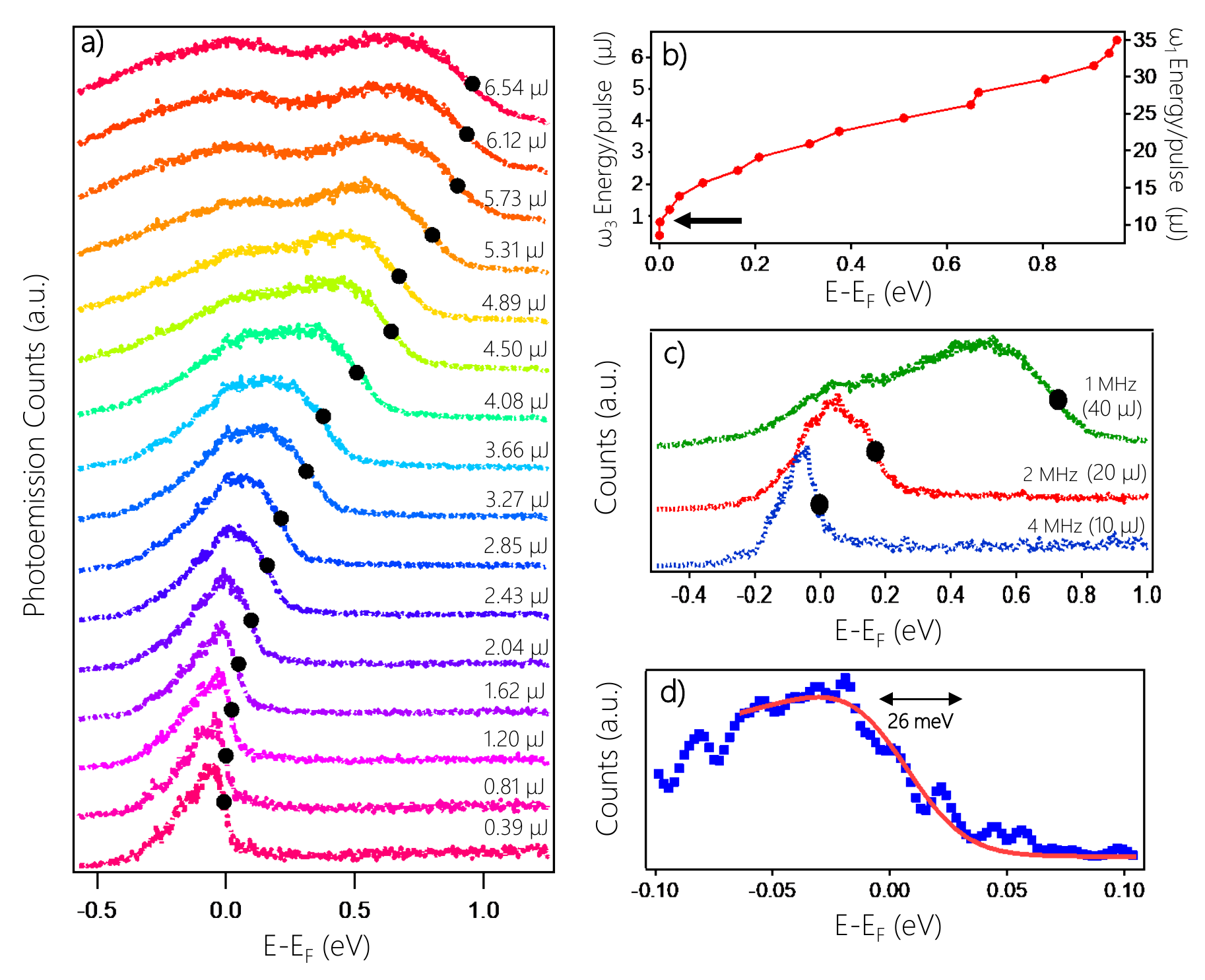}
\caption{\label{fig:Space_Charge}\textbf{Space charge effect.} Panel (\textbf{a}) shows EDCs extracted from photoemission intensity spectra $I(E,k_x)$ of polycrystalline gold. The curves are normalized and shifted for clarity. The graph highlights the evolution of the Fermi edge position (marked with black dots) at different pulse energies. The values of $E_F$ are reported in panel (\textbf{b}) as a function of the $\omega_3$ (left axis) and $\omega_1$ (right axis) energy per pulse. The space charge effect responsible of the Fermi edge shift disappears when only the 10\% of the maximum third harmonic power is used. It corresponds to $\sim0.8$ $\mu$J/pulse for the third harmonic and $\sim10\mu$J/pulse for the fundamental at 1035 nm. (\textbf{c}) Space charge can be mitigated increasing the repetition rate at constant power. At 1 and 2 MHz the distortion effect is still present, at 4 MHz, instead, the measurement is space-charge-free. In panel (\textbf{d}) we used the blue curve in panel (c), zoomed out in the energy range $\pm 100$ meV, to extract the energy resolution of the apparatus. The red line is the fit to the data (see text). The resulting net energy resolution is $\simeq 26$ meV.}
\end{figure}

One of the well-known problems of photoemission experiments using pulsed laser sources is the space charge effect. The use of intense laser pulses, with a very short time duration, causes a large number of electrons being extracted from the material and occupying a small volume. These electrons in a crowded cloud experience strong Coulomb repulsions which modifies both the longitudinal (energy) and lateral (momentum) distributions. As a consequence, the Fermi edge is shifted to higher energies and the revealed signal is severely distorted with respect to the original band structure \cite{Hellmann2009}. We evaluated the space charge effect in our system on a sample of polycristalline gold in order to avoid significant spectral features other than the Fermi edge. Energy distribution curves (EDCs) integrated over $10^{\circ}$ in the momentum direction are reported in Figure \ref{fig:Space_Charge}(a). The top curve, taken at 1 MHz, shows a big distortion of the Fermi edge that appears shifted by $\sim 1$ eV from the correct position ($E-E_F=0$) when the driving light at 345 nm is kept at 80\% of its maximum power, corresponding to an energy/pulse of 6.54 $\mu$J. One obvious solution to mitigate the space charge distortion is to reduce the pulse intensity. The Coherent Monaco has a built-in attenuator that allows a continuous remote control of the pulse intensity. Exploiting this feature, while monitoring the 345 nm energy/pulse, we gradually reduced the pulse intensity and we followed the evolution of the Fermi edge shift as reported in Figure \ref{fig:Space_Charge}(a) (the curves have been opportunely normalized and shifted for clarity). The position of the Fermi energy (black dots) has been extracted from a Fermi-Dirac distribution fit convoluted with a Lorentzian lineshape and a gaussian, accounting for experimental resolution, and then reported in the panel (b) of Figure \ref{fig:Space_Charge} as a function of the third harmonic energy/pulse (left axis) and the respective fundamental energy/pulse (right axis). From the graph we can identify a region below $\sim0.81$ $\mu$J/pulse (10\% of the third harmonic maximum power, highlighted by the arrow) in which the spectra are space-charge-free. The consequent reduction in the counts rate is compensated by the base operational repetition rate of 1 MHz that assures fast acquisition times. The space-charge-free condition at 1 MHz r.r. utilizes only $\sim 25\%$ (10.56 $\mu$J/pulse) of the total output power at 1035 nm leaving available a lot of power for the pump path, opening the possibility to efficiently seed an Optical Parametric Amplifier for a tunable pump beam, contributing to the versatility of the whole setup.
The laser system also allows to tune the repetition rate in excess of 1 MHz (up to 50 MHz), at the expense of the available energy/pulse, which can be calculated from the constant output power of 40 W in the 1-50 MHz r.r. range. This feature plays an important role in photoemission experiments. In fact, even though the direct reduction of the pulse intensity (filters, spoiled harmonic efficiency etc.) is efficient in the space charge reduction, it impoverishes the counts statistic since less electrons per second are emitted. An alternative is to keep constant the output power of the laser, and increase the repetition rate: the energy for each pulse is reduced, simultaneously controlling the space charge, but the number of pulses is larger. With this approach, in Figure \ref{fig:Space_Charge}(c), we reported the extracted EDCs keeping maximum the output power of the laser and changing the repetition rate from 1 MHz (standard operation with 40 $\mu$J/pulse) to 2 MHz (20 $\mu$J/pulse) and 4 MHz (10 $\mu$J/pulse). After a normalization we can see that at 2 MHz a space charge effect is still present while it is completely vanished at 4 MHz. This fixes the potentiality of our system that, thanks to the easy control on the pulse intensity and on the repetition rate, can perform time resolved ARPES measurements with a VUV 10.8 eV probe with an unprecedented repetition rate, throughput, versatility and reliability.
From the fit of the EDC curve measured at 4 MHz and reported in Figure \ref{fig:Space_Charge}(d) on a $\pm 100$ energy range, we extracted an energy resolution of $\simeq 26$ meV which is the usual resolution of our analyzer previously measured. This assure that the intrinsic energy bandwidth of the 10.8 eV light is not impacting the ARPES measurements.

\section{\label{sec:level1}Case studies: TR-ARPES on B\lowercase{i}$_2$S\lowercase{e}$_3$ topological insulator and WT\lowercase{e}$_2$ semimetal}
\subsection{\label{sec:level2}Bi$_2$Se$_3$ topological insulator}
\begin{figure}
\includegraphics[width=\linewidth]{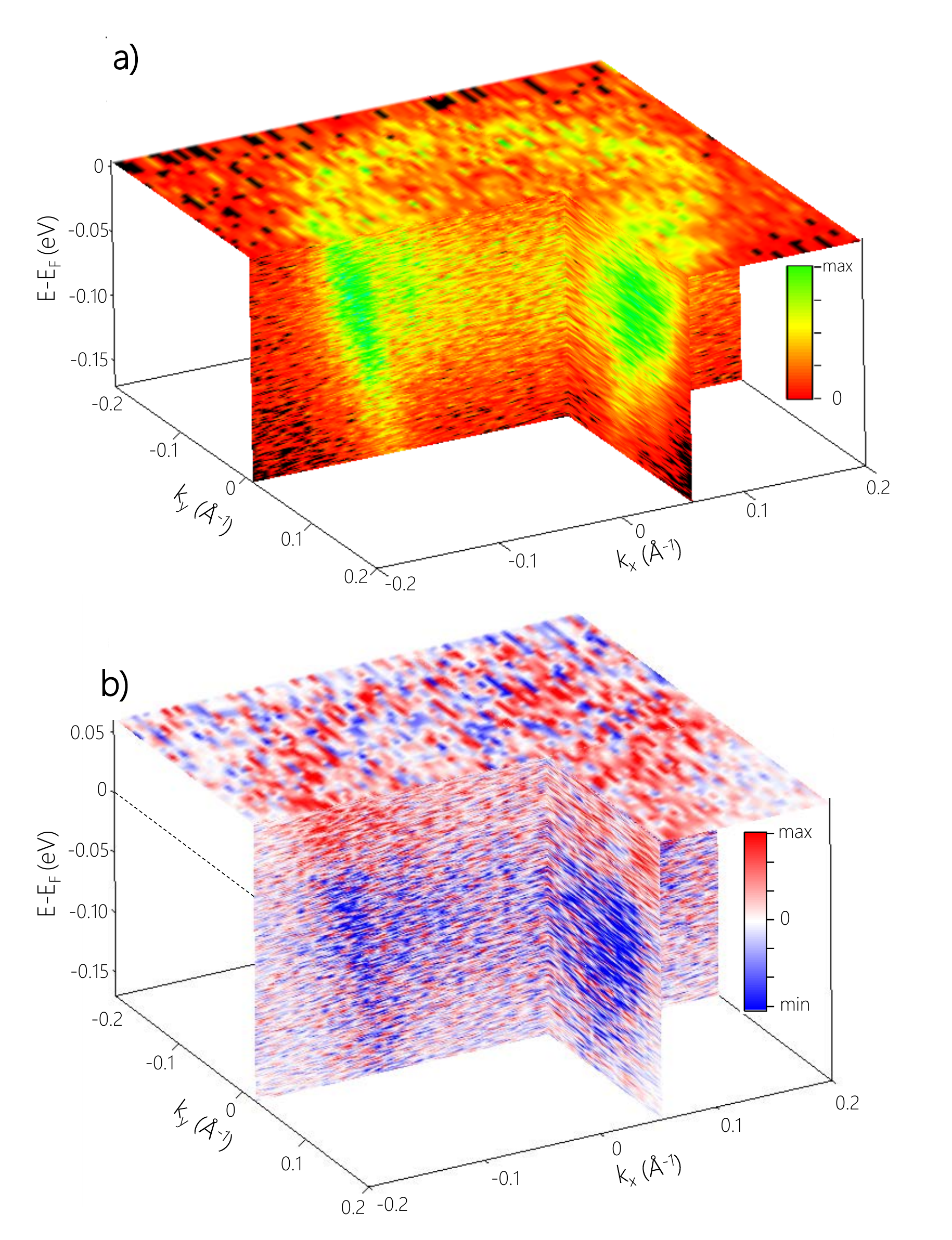}
\caption{\label{fig:Cube}\textbf{3-dimensional dispersion ($E,k_x,k_y$) of Bi$_2$Se$_3$}. Panel (\textbf{a}) shows the equilibrium 3-dimensional dispersion of the Bi$_2$Se$_3$. The top view of the map shows the photoemission intensity of the Fermi surface with its peculiar circular shape. In the $(E_B,k_x)$ slice it is possible to identify the three bands of the Bi$_2$Se$_3$ (also sketched in Figure \ref{fig:Time_Resolved2}): the linear development of the TSS in the outer part, the diffuse intensity of the BCB in the inner part and the two-dimensional EG as the more intense rim of the conduction band. In panel (\textbf{b}) we reported the differential map obtained by subtraction of the equilibrium map to the one resulting after photoexcitation at 1.2 eV. The top view shows the constant energy surface 50 meV above $E_F$. The differential map is taken at $t=1$ ps, where the signal is found to be maximum.}
\end{figure}

\begin{figure*}
\includegraphics[width=\linewidth]{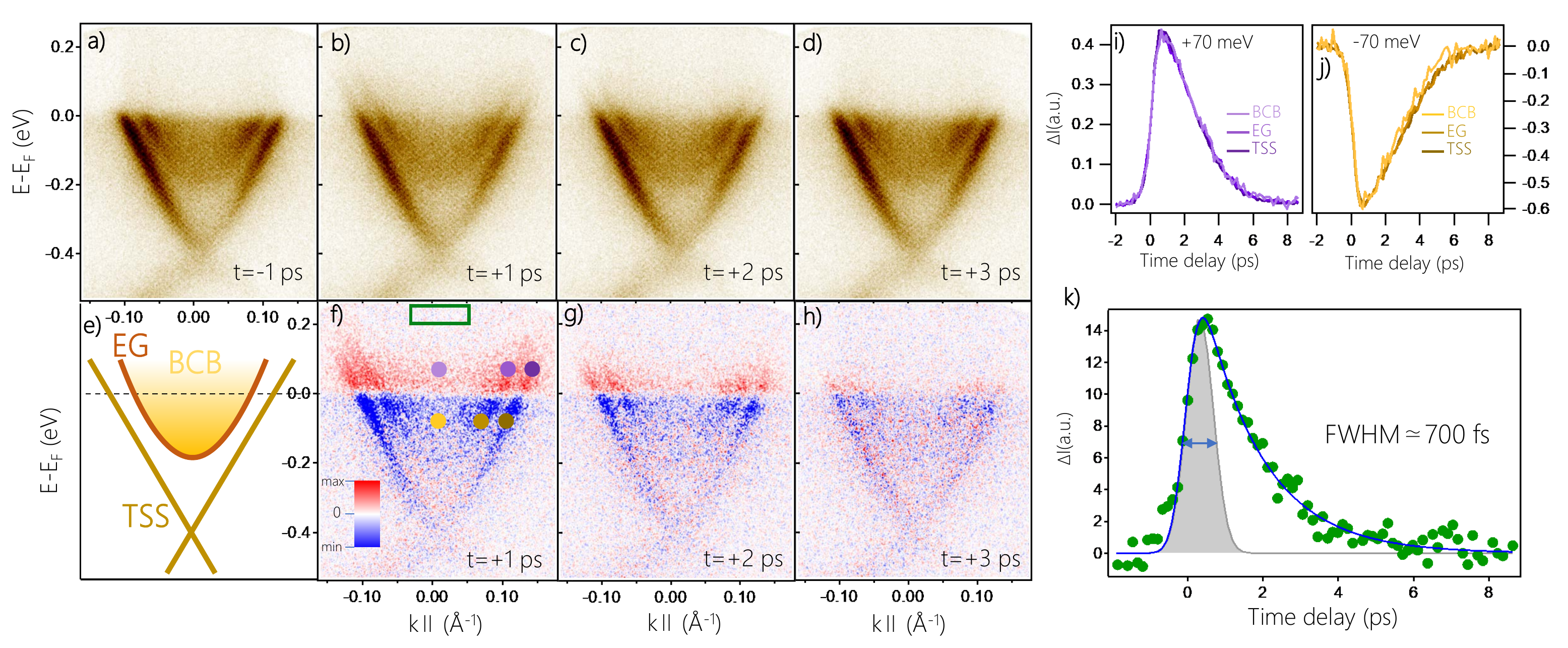}
\caption{\label{fig:Time_Resolved2}\textbf{Dynamics of a 2-dimensional ($E_B,k_x$) map of Bi$_2$Se$_3$.} The top strip shows the $E$ vs $k_x$ dispersion at $k_y=0$ before the arrival of the pump (\textbf{a}) and 1 ps, 2 ps, 3 ps (\textbf{b}, \textbf{c}, \textbf{d}) after photoexcitation with 1.2 eV photons. Panel (\textbf{e}) shows the three spectral features reported in literature \cite{Bianchi2010}: the topological surface state (TSS), the bulk conduction band (BCB) and the two-dimensional surface electron gas (EG). Photoinduced spectral modifications are more directly visible in the differential maps (\textbf{f}-\textbf{h}), obtained by subtraction of the equilibrium map to the maps in (b-d). Full dynamics have been extracted 70 meV above (purple shades) and below (yellow shades) the Fermi level, integrating the spectra in the regions indicated by the coloured areas in panel (f). These dynamics are superimposed in panels (\textbf{i}) and (\textbf{j}), displaying no appreciable differentiation. Panel (\textbf{k}) shows the time trace obtained by integrating over the green rectangle in panel (f). This trace has been fitted to a single exponential convoluted with a gaussian accounting for the experimental time resolution; the gaussian FWHM extracted from this fit is $\sim700$ fs.}
\end{figure*}

To prove the functionality and the reliability of our system we performed TR-ARPES measurements on the bismuth selenide topological insulator Bi$_2$Se$_3$. Thanks to peculiar surface states in which, because of the chiral spin structure, backscattering is not allowed, topological insulators have been widely studied and proposed as candidates for spintronics and quantum computing applications \cite{Garate2010}, where the spin coherence is crucial. Like the whole family of the topological insulators \cite{Zhang2009}, this material is characterized by an energy gap between the occupied and unoccupied states in the bulk band structure and by a gapless topological surface state (TSS) at the surface \cite{Pan2011, Lu2012}. In the electronic band structure, the TSS inside the band gap has a linear dispersion giving origin to a Dirac cone with a Dirac point at a binding energy of $\sim 0.3-0.4$ eV below the Fermi energy \cite{Zhu2011}. In grown Bi$_2$Se$_3$ crystals, the impurities due to Se vacancies lead to a net n-doping of the material with a consequent partial population of the bulk conduction band (BCB) crossing the Fermi energy. A sketch of the Bi$_2$Se$_3$ electronic band structure is shown in Figure \ref{fig:Time_Resolved2}(e). For a comprehensive understanding of the electronic band structure of materials, a 3-dimension $I(E_B, k_x, k_y)$ photoemission intensity map is required. Standard hemispherical analyzers can measure 2-dimension intensity maps $I(E_B,k_x)$ at a single $k_y$ fixed by the geometry of the experiment, while a 3-dimension map is obtained by stitching of the single slices taken at different values of $k_y$. 
The characteristics of our ARPES setup at 10.8 eV, which combines a good beam focusing and a very high throughput thanks to the MHz r.r., along with the six degrees of freedom of the manipulator, can be exploited to measure the complete Fermi surface of most of the materials both at the equilibrium and out-of-equilibrium. In Figure \ref{fig:Cube}(a) we measured the complete photoemission intensity I($E_B$,$k_x$, $k_y$) as a function of binding energy $E_B$ and parallel momenta $k_x$ and $k_y$. This is obtained by stitching together single $E_B$-$k_x$ spectra taken at different tilt angles $\phi$. We span over a range of $\phi=\pm 20^{\circ}$ around the normal emission with a step of $0.5^{\circ}$ and the sample oriented along the $\overline{\rm \Gamma}$-$\overline{\rm M}$ direction. Both the momentum directions have an extension of $\pm0.2$ \AA$^{-1}$ while the energy window is set from $-0.17$ meV to $+0.05$ meV. All the measurements shown in the present work were taken at a temperature $T=110$ K. 
We show the photoemission intensity at the Fermi level $E-E_F=0$ and the cut along the $\overline{\rm \Gamma}$-$\overline{\rm M}$ direction at $k_y=0$ where it is clearly visible the presence of both the TSS band and the BCB band. Figure \ref{fig:Cube}(b) represents the maximum photoemission intensity variation, recorded at t=1 ps, as induced by a 30 $\mu$J/cm$^2$ beam at 1035 nm. Both the pump and the probe are $p$ polarized. The excitation at 1.2 eV drives a direct transition from the valence band to high-lying unoccupied bulk and surface states, and after $\sim1$ ps these electrons scatter down to the TSS and BCB \cite{Sobota2014}. The top cut shows the shape of the constant-energy surface 60 meV above the Fermi energy to allow a complete understanding of the electronic band structure across the Fermi level.

The study of the overall relaxation dynamics was carried out on a single $E_B$-$k_x$ plane, precisely at $k_y=0$ that identifies the $\overline{\rm \Gamma}$-$\overline{\rm M}$ high symmetry direction. The binding energy window was set to $-0.50 < E_B < 0.25$ eV. The fluence and polarization of the beams are the same as in Figure \ref{fig:Cube}(b). The acquisition time was instead increased to obtain a better signal-to-noise ratio and a better quality of the ARPES spectra. In the top strip of Figure \ref{fig:Time_Resolved2} we show four ARPES spectra taken at different delays of the probe pulse. Panel (a) shows the ARPES spectrum at equilibrium, before the pump excitation. All the features of Bi$_2$Se$_3$ down to 500 meV below$E_F$ are clearly visible, with the linear dispersion of the TSS which originates the Dirac point at $\sim 400$ meV below $E_F$. Inside the Dirac cone there is a diffuse intensity due to the population of the bottom of the BCB having its minimum at $E_B=200$ meV. All around the BCB there is an intense and narrow rim which is assigned to a quantum-confined two-dimensional electron gas (EG) formed near the sample surface, as already reported in \cite{Bianchi2010}. This is a consequence of the aging and the deterioration of the surface sample \cite{Chen2012} after the cleave and is driven by the creation of extrinsic defects or impurities at the surface. In panels (b),(c) and (d) we show the same spectra respectively 1 ps, 2 ps and 3 ps after the pump excitation. Here we can appreciate the photoinduced spectral changes: states above $E_F$ are populated by depletion of the states below $E_F$. For a more direct visualization of the effect of the pump excitation, we reported in panel (f), (g) and (h), the difference between the excited spectra with the spectrum at negative delay. Thanks to the high momentum resolution, the contributions from each of the three bands under investigation can be clearly distinguished.

The power of the time-resolved ARPES technique is to unveil the band structure in the unoccupied region of the spectrum (otherwise inaccessible) and to reveal the relaxation dynamics of the carriers inside each band. The possibility to disentangle different interband or intraband thermalization channels in a TR-ARPES experiment can be applied to the situation of Bi$_2$Se$_3$, in which the physical nature of the topological state and the two-dimensional electron gas is very different (since the latter is fully spin-degenerate, while the former is not). The same holds for the bulk conduction band which has a 3D nature while both the EG and the TSS exist only at the surface. Many time-resolved studies have been carried out on the band dynamics of Bi$_2$Se$_3$, all of them concentrating on the differences between the TTS and the CB \cite{Sobota2014, Hajlaoui2012, Crepaldi2012, Wang2012}, but information on the dynamics of the two-dimensional electron gas feature is still lacking. In Figure \ref{fig:Time_Resolved2}(i) and (j) we show the extracted relative photoemission intensity variation $\Delta I$ as a function of the pump-probe delay $t$ in the regions indicated by coloured circles in panel (f). We investigated the relaxation dynamics of the three different bands 70 meV above (shades of purple) and below (shades of yellow) the Fermi level. We superimposed the three curves (after intensity renormalization) and conclude that no differences are evident. From previous studies it has been proved that, after excitation, the interband scattering occurs in the first 200 fs. In this time window, at similar temperatures, there is a transfer of spectral weight from the BCB to the TTS, mainly due to a phonon scattering mechanism, which reflects the scattering of the electrons between the surface states and the bulk conduction band \cite{Wang2012}. At longer delays the cooling mechanism is mainly due to intraband scattering, thus being very similar for the three different bands \cite{Hajlaoui2012}. Since our time resolution exceeds the timescale of the fastest interband scattering process, no differences can be appreciate in the relaxation dynamics of the TTS, BCB and EG bands.

\begin{figure*}
\includegraphics[width=0.9\linewidth]{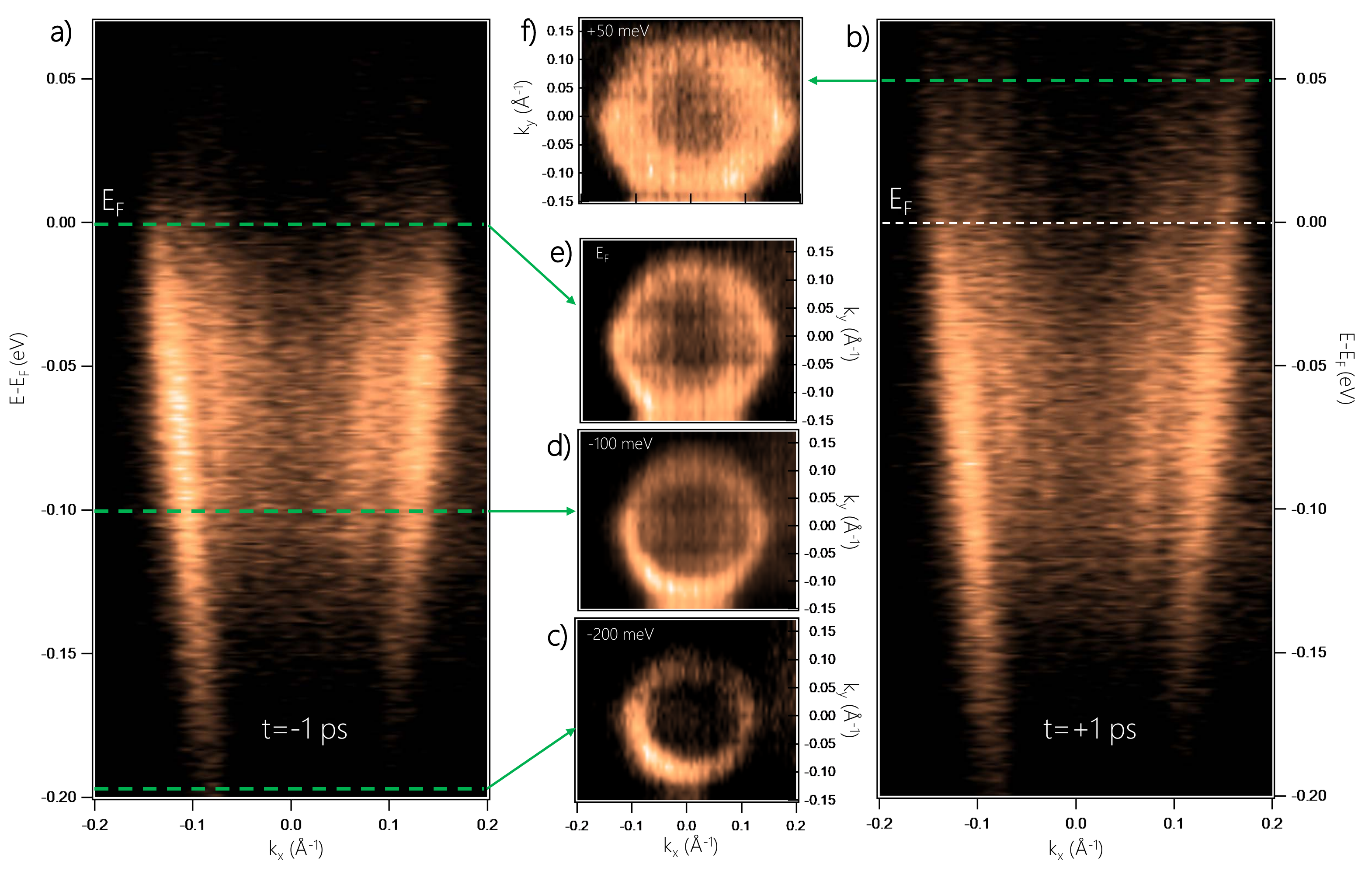}
\caption{\label{fig:Fermi_Surface}\textbf{Evolution of the Fermi surface shape of Bi$_2$Se$_3$ at different binding energies.} Panel (\textbf{a}) and (\textbf{b}) show two-dimensional maps taken at $k_y=0$, respectively before ($t=-1$ ps) and after ($t=+1$ ps) the optical excitation at 1.2 eV. Constant energy maps (\textbf{c}-\textbf{f}) are taken at the energies indicated by the green lines (200 and  100 meV below the Fermi level, at the Fermi level, and 50 meV above, respectively). The TSS band evolves from circular to hexagonal shape while approaching the Fermi level. This effect is more striking above the Fermi level, where the hexagonal symmetry is broken along the $k_x$ (here horizontal) direction.}
\end{figure*}

Besides the analysis of the physical aspects of the relaxation dynamics of the Bi$_2$Se$_3$, we used the same measurement to estimate the overall time resolution of the system. Indeed, a direct measurement of the probe time duration is not achievable. We extracted the relaxation dynamics of the electronic population far from the Fermi level to reduce the effects of the population dynamics \cite{Crepaldi2012} and to approach a purely exponential decay. For these reasons, we integrated over the area enclosed in the green rectangle in Figure \ref{fig:Time_Resolved2}(f) at $E-E_F\simeq0.22$ eV above the Fermi level and we reported the relaxation curve in Figure \ref{fig:Time_Resolved2}(k). The fit function is the convolution of a gaussian function, accounting for the experimental resolution, and a single exponential function. The fitting procedure revealed a decay time $\tau=1.57$ ps and an overall time resolution of $\sim 700$ fs.
\\
\\
A careful analysis of the full $I(E,k_x,k_y,t)$ map reveals an interesting behaviour of the shape of the Fermi surface at different binding energies. Due to the large band gap ($\sim 0.3-0.4$ eV) Bi$_2$Se$_3$ approaches the features of an ideal Dirac cone, with a perfectly circular Fermi surface. However, as noted by previous studies \cite{Chen2012, Bianchi2010, Kuroda2010}, the hybridization of the surface states with the bulk states leads to a deformation of the circular symmetry. This deformation occurs more prominently at lower binding energies, and bears the hexagonal symmetry of the lattice since it is shared also by the BCB. In Figure \ref{fig:Fermi_Surface}, we show the 2-dimension map $I(E_B,k_x)$ cut at $k_y=0$ of the 3-dimensional map of Figure \ref{fig:Cube} before the pump excitation (panel (a)) and 1 ps after the perturbation (panel (b)). Constant energy cuts $I(k_x,k_y)$ have been taken at binding energies marked by dashed green horizontal lines, integrating the spectra in an energy window of 30 meV. From the equilibrium map we took cuts at $E_F$ and below $E_F$ (0.2 eV and 0.1 eV), while from the excited map we took a cut above $E_F$ (0.05 eV). These are shown in the central vertical strip in panel (e), (c), (d) and (f) respectively. The evolution of the hexagonal deformation while approaching the Fermi level can be clearly seen in these images. The same holds for the shape of the inner EG band, although less visible. Photoexcitation allows to probe the shape of the Fermi surface also above the Fermi level (panel (f)). Here we detect a sharp hexagonal shape which covers almost the whole momentum space under investigation. However, the symmetry of the hexagon is deformed along the $k_x$ axis (here, the horizontal one) . It is worth to mention that the linear polarization of the pump was set to be parallel to the $k_x$ direction. This suggests that the origin of the deformation effect might be attributed to the polarization of the optical excitation. This opens a new space for interesting future investigations.

\subsection{\label{sec:level3}WTe$_2$ semimetal}
\begin{figure}
\includegraphics[width=\linewidth]{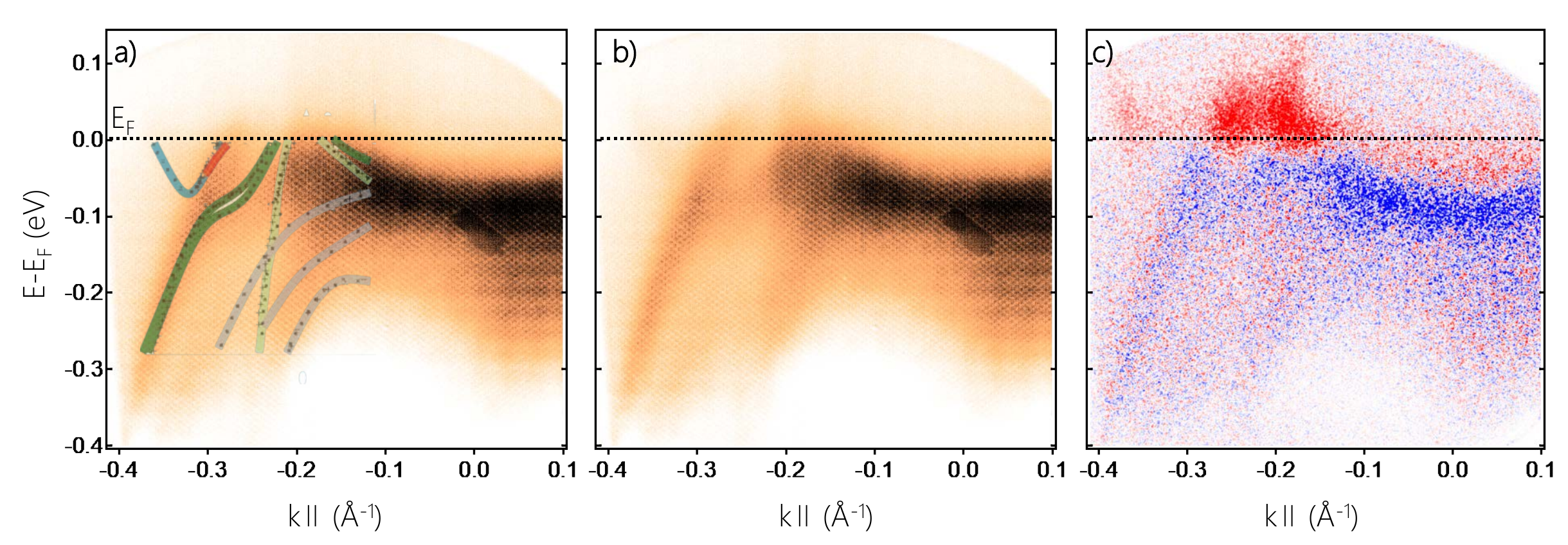}
\caption{\label{fig:WTe2}\textbf{TR-ARPES experiment on WTe$_2$.} (\textbf{a}) Measurement of the photoemission intensity $I(E,k_x)$ at equilibrium, at $k_y=0$. The lines (from \cite{Bruno2016}) are a guide for the eye to the main features of the WTe$_2$ band structure. The blue line identifies the electron pockets while the green ones refers to the hole pocket. The high photoemission intensity on the right branch of the electron pocket can be ascribed to the presence of a surface state (red line). Panel (\textbf{b}) shows the same map at 1 ps after an optical excitation at 1.2 eV photon energy. Panel (\textbf{c}) shows the difference between the excited and the equilibrium states.}
\end{figure}

We measure the non-equilibrium band structure of WTe$_2$, a two-dimensional transition metal  dichalcogenide. It is a new type of Weyl semimetal with strongly tilted Weyl cones \cite{Wang2016}. WTe$_2$ has become popular for its non-saturating magnetoresistance \cite{Ali2014} and the discovery of pressure induced superconductivity \cite{Kang2015}. The Fermi surface of this material is characterized by the presence of electron and hole pockets in the $(E,k_x)$ plane along the $\overline{\rm \Gamma}$-$\overline{\rm X}$ direction as depicted in Figure \ref{fig:WTe2}(a). The green lines represents the bulk bands responsible for the electron pockets while the blue ones for the hole pockets. The red line identifies a surface state Fermi arc emerging out of the bulk electron pocket in its immediate vicinity \cite{Bruno2016}. In Figure \ref{fig:WTe2}(b) we show the band structure after excitation by a 1.2 eV optical pulse. The promotion of the electrons along the two branches of the electron pocket is induced and probed. The right branch displays a higher intensity due to the contribution of the surface state. The hole pockets bands rejoin just above the Fermi level as predicted by density functional theory calculations \cite{Bruno2016}. For a better visualization of the effect of the pump excitation on the WTe$_2$ bands, we reported in Figure \ref{fig:WTe2}(c) the differential map obtained subtracting the ARPES intensity at negative delay from the pumped one. This measurement proves that the photon energy of our experimental setup allows to reach large momenta while maintaining a small tilt angle of the sample with great benefits on the quality of the measurements.

\section{\label{sec:level4} Discussion and Conclusions}
In this work we describe an experimental system for time-resolved ARPES with probe photon energy of 10.8 eV, $<26$ meV energy resolution, $\sim 700$ fs time resolution and a repetition rate up to 4 MHz. The development here presented overcomes the momentum mapping capability achievable with standard non linear crystal-based setups and the low conversion efficiency guaranteed by HHG systems. The VUV light is generated, in phase matching condition, as a third harmonic generation process in Xenon, with an overall conversion efficiency of $\sim 10^{-4}$-$10^{-5}$. The favourable phase matching condition is obtained by using a Yb fiber-based laser producing pulses at 1035 nm. Its ninth harmonic falls in the negatively dispersive region of the Xenon gas (113.5-117 nm). The tunability of the pulse intensity and of the repetition rate assures a high versatility of the system. We proved that very high throughput measurements at 4 MHz, i.e. 10 $\mu$J pulse energy, can be performed completely avoiding distortions caused by space-charge effects. This configuration can be used especially for equilibrium measurements, since the thermal heating provoked by the pump excitation could impact the sample under scrutiny. Alternatively, lower repetition rate can be used for time-resolved measurements. We proved that, working at 1 MHz repetition rate, a space charge free condition is achieved using an energy/pulse of only 10 $\mu$J ($\sim25\%$ of the total laser output power), thus leaving the excess power available for pumping OPA and other experimental purposes. Besides the high conversion efficiency, the physical process for the VUV generation permits to have a complete control of the beam focusing. The average of 260 $\mu$m as FWHM of the spot size on the sample makes this system suitable for TR-ARPES also on small samples. Moreover, a control of the linear polarization of the VUV light is also available by employing a $\lambda/2$ waveplate on the driving beam. All these characteristics, along with the six degrees of freedom of the manipulator, allows to measure the out-of-equilibrium photoemission intensity $I(E,k_x,k_y,t)$ of the complete first Brillouin zone of most materials placing our system to the frontier of the TR-ARPES experiments in condensed matter physics. We proved the capabilities of our TR-ARPES setup on the prototypical topological insulator Bi$_2$Se$_3$. We have shown how the high energy and momentum resolutions can bring out new interesting and unexplored results on this material, like the relaxation dynamics of the two-dimensional electron gas on its surface or the shaping evolution of the Fermi surface of the unoccupied states above the Fermi level. Similar measurements have been performed on a WTe$_2$ sample to study the behaviour of its electronic bands far from the $\Gamma$ point. The accessibility of such larger momenta are a peculiar feature of the experimental setup presented in this work. The T-ReX laboratory at Elettra Sincrotrone in Trieste has now its third laser-based UV-VUV source for TR-ARPES measurements. In fact, along with the 10.8 eV photons it hosts a traditional 6.2 eV setup and a tunable 17-31 eV HHG system. The three experimental setups, characterized by diverse specifications and seeded by different laser systems, are remarkably driven in the same photoemission chamber. The capability of easily switching among these sources, alternatively privileging photon energy, energy resolution and time resolution makes this facility unique on the world scene.

\begin{acknowledgments}
The authors would like to thank Aleksander De Luisa for the useful suggestions on the design of the experiment and for the realization of the technical drawings of the vacuum refocusing chamber.
\end{acknowledgments}

\bibliography{Peli_11eV}

\end{document}